\documentclass[a4paper,
               ]{jacow}
\usepackage{pdfpages,multirow,ragged2e,booktabs} %
\usepackage{url} 
\usepackage{hyperref} 
\usepackage{textgreek} 
\makeatletter%
	\ifboolexpr{bool{xetex}}
	 {\renewcommand{\Gin@extensions}{.pdf,%
	                    .png,.jpg,.bmp,.pict,.tif,.psd,.mac,.sga,.tga,.gif,%
	                    .eps,.ps,%
	                    }}{}
\makeatother

\ifboolexpr{bool{xetex} or bool{luatex}} 
 {}                                      
 {\usepackage[utf8]{inputenc}}           

\usepackage[USenglish]{babel}

\ifboolexpr{bool{jacowbiblatex}}%
 {%
  \addbibresource{jacow-test.bib}
  \addbibresource{biblatex-examples.bib}
 }{}
\begin{document} 
\title{AI-driven neutrino diagnostics and radiation-hard beam instrumentation for next-generation neutrino experiments\thanks{}}
\author{S. Ganguly\thanks{sganguly@fnal.gov}, Fermi National Accelerator Laboratory, Batavia, USA} 
	
\maketitle

\begin{abstract}
   The Long Baseline Neutrino Facility (LBNF) at Fermilab will deliver a high-intensity, multi-megawatt neutrino beam to the Deep Underground Neutrino Experiment (DUNE), enabling precision tests of the three-neutrino paradigm, CP violation searches, neutrino mass ordering determination, and supernova neutrino studies. In order to accelerate DUNE's physics reach and ensure robust beam operations, we propose an integrated AI-driven framework with real-time diagnostics and radiation-hardened instrumentation. A physics-informed digital twin is at the heart of this Real-Time Beam Integrity Monitor. By reconstructing pion phase space from muon profiles and exploiting magnetic horn optic linearity, it enables spill-by-spill beam correction and flux stabilization. By using this approach, flux-related systematics could be reduced from 5\% to 1\%, potentially accelerating the discovery of CP violations by four to six years. Complementing this, a US–Japan R\&D effort will deploy a LGAD-based muon monitor in the NuMI beamline. Time of Flight (ToF) measurements can be acquired with picosecond precision using this radiation-hard system, enhancing sensitivity to horn chromatic effects. Simulations confirm strong responses to these effects. 
   Machine learning models can predict beam quality and horn current with sub-percent accuracy. This scalable, AI-enabled strategy improves beam fidelity and reduces systematics, transforming high-power accelerator operations.  
\end{abstract}

\section{Introduction}

The Long Baseline Neutrino Facility (LBNF)~\cite{papadimitriou:ipac16-tupmr025} at Fermilab will deliver a multi-megawatt, high-intensity neutrino beam to the massive detectors of the Deep Underground Neutrino Experiment (DUNE)~\cite{papadimitriou:ipac18-tupaf075}, enabling precise tests of the three-neutrino flavor paradigm, direct searches for CP violation, determination of neutrino mass ordering, and detailed studies of supernova neutrino interactions. Although DUNE is baselined to achieve its physics goals independently, recent advancements in AI-driven diagnostics and real-time accelerator control provide a unique opportunity to surpass baseline performance at a significantly lower cost.

A major driver of sensitivity in DUNE is the precision of the neutrino flux prediction. Achieving percent-level accuracy in flux normalization can reduce the overall systematics to below 3\%, potentially shortening the runtime needed for key measurements by several years. Current beam flux uncertainties at the source are approximately 2-3\%, and new approaches are required to suppress this contribution further. 

\begin{figure}[htbp]
    \centering
    \includegraphics[width=0.75\columnwidth]{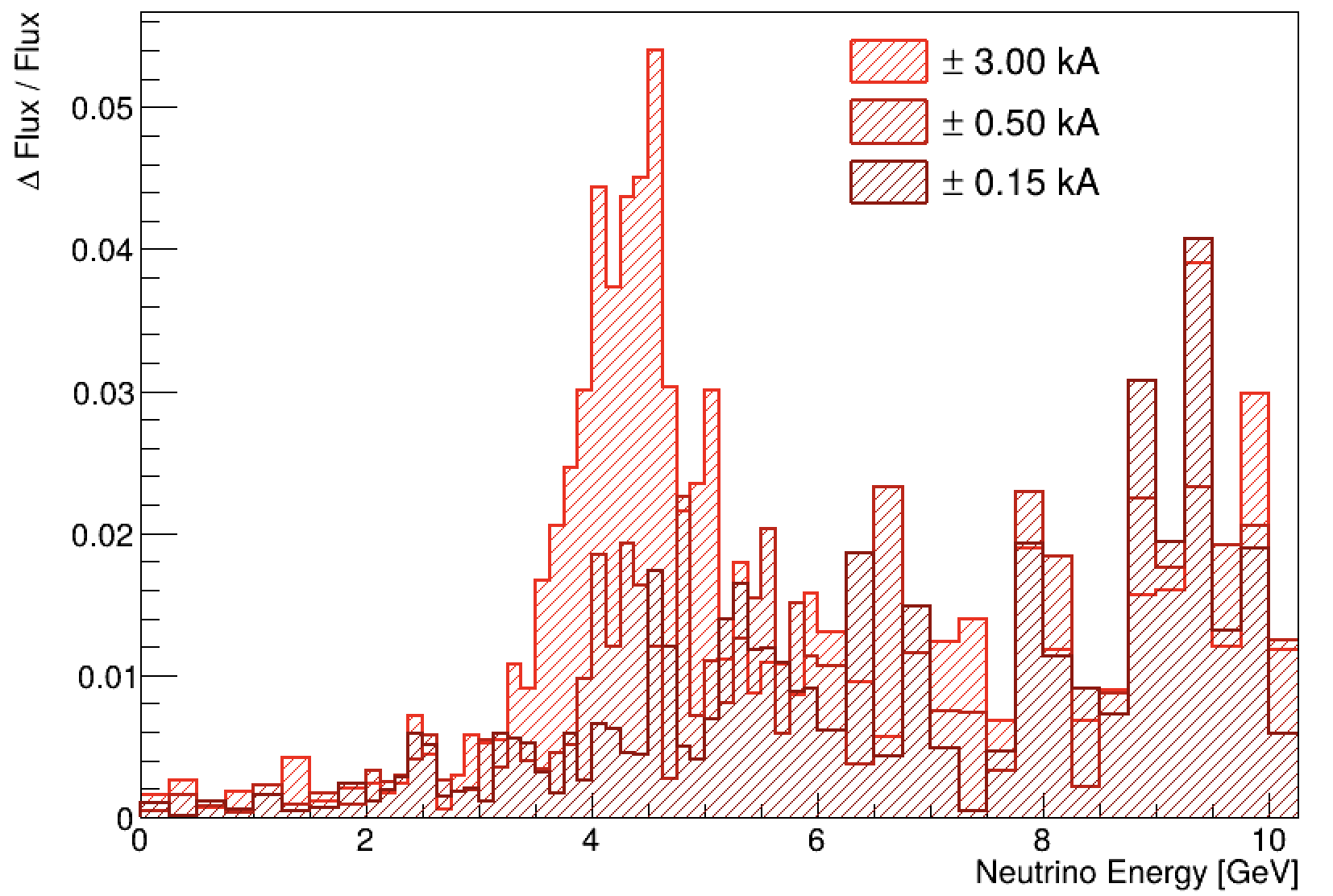} 
    \caption{Relative change in neutrino flux as a function of energy for horn current variations of \(\pm3\), \(\pm0.5\), and \(\pm0.15\)~kA. Smaller uncertainty in horn current reduces the fractional flux variation by up to a factor of five.}
    \label{fig:horn_uncertainty_flux}
\end{figure}

Muon monitors, placed downstream of the decay pipe, offer a unique window into the phase space of parent pions. Since muons and neutrinos are produced together in pion decays, measurements from the muon monitors provide indirect but valuable information about the resulting neutrino flux. The current implementations, however, are limited in time resolution and cannot capture correlations between events.
A machine-learning-based diagnostic framework aims to improve flux prediction and enable adaptive beam control in this work. 
 The approach uses correlations between real-time signals from muon monitors and upstream instrumentation to infer the pion phase space and chromatic beamline effects ~\cite{q6l6-wywy}, such as those arising from horn current fluctuations. These models are trained using a physics-informed digital twin, allowing us to bridge simulation and real-world conditions. 
 Figure~\ref{fig:horn_uncertainty_flux} demonstrates the impact of horn current uncertainty on neutrino flux prediction. A typical uncertainty of \(\pm3\)~kA (1\%) leads to flux variations exceeding 5\% in the peak energy region, while reducing the uncertainty to \(\pm0.15\)~kA suppresses this effect to below 1\%. This motivates the need for precision monitoring and inference frameworks that can resolve such variations on a spill-by-spill basis. The machine learning strategy developed in this work aims to achieve this resolution using only muon monitor signals, enabling both tighter flux constraints and future feedback-based beam optimization.

\section{Muon Monitoring and Beam Correlation}
\begin{figure}[htbp]
    \centering
    \includegraphics[width=0.95\columnwidth]{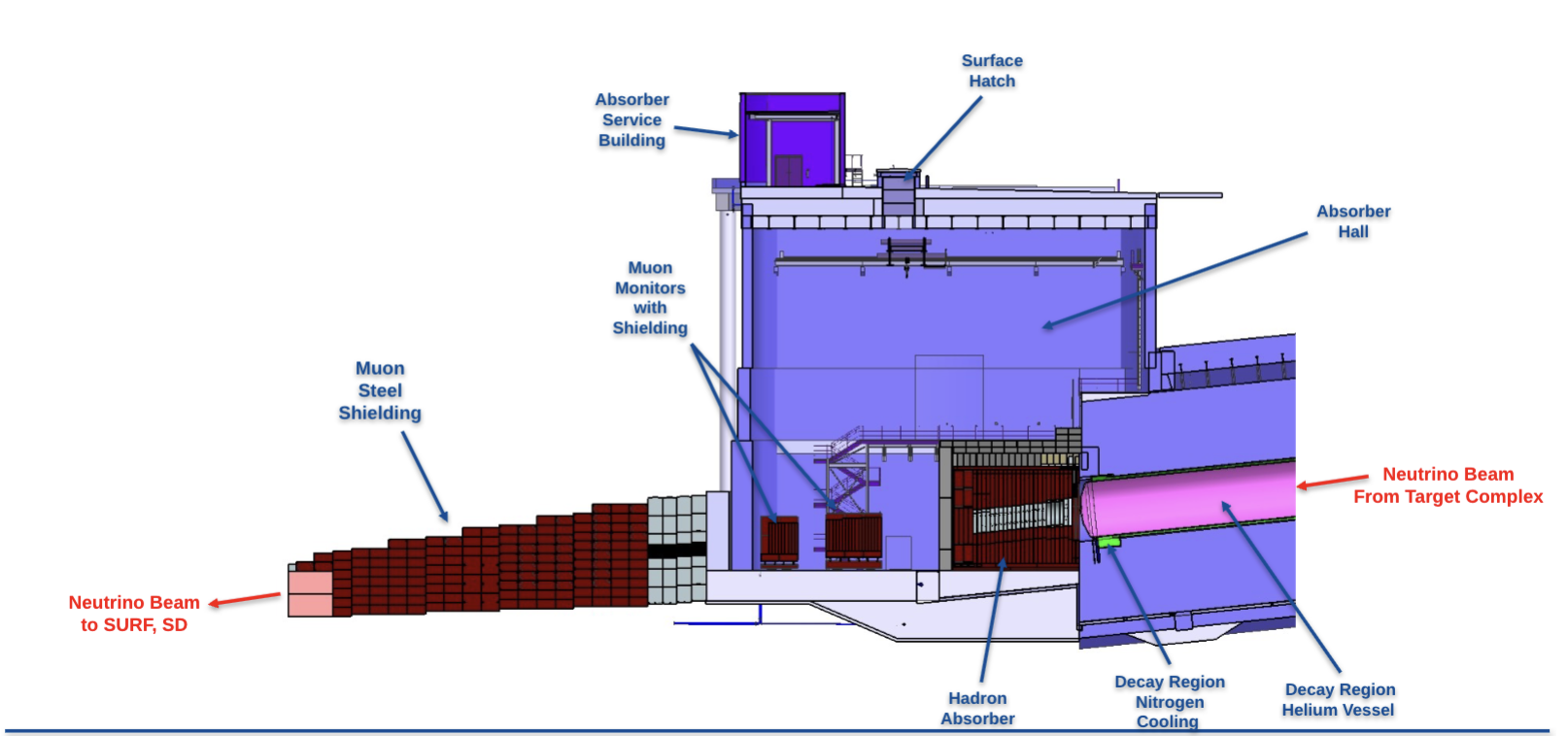} 
    \caption{Cross-sectional schematic of the LBNF absorber and downstream monitoring region. 
    Muon monitors are placed in three shielded alcoves downstream of the Hadron Absorber. These locations offer energy discrimination via natural pion decay kinematics.}
    \label{fig:absorber_layout}
\end{figure}

As part of the Neutrino Beam Instrumentation system for LBNF, dedicated detectors monitor secondary and tertiary particles. This provides essential information for beam alignment, anomaly detection, and long-term diagnostics.
Its two most important components are the Hadron Alignment Detector System (HaDeS) and Muon Monitor System (MUMS).
They serve as the main interface between upstream beamline parameters and downstream neutrino flux behavior. The spatial configuration of these systems is illustrated in Fig.~\ref{fig:absorber_layout}, showing the placement of muon monitors within shielded alcoves behind the hadron absorber.
HaDeS is derived from the NuMI hadron monitor design~\cite{Zwaska_2006}, which utilizes a $7 \times 7$ array of parallel-plate ionization chambers, covering a total area of 1~m$^2$. For LBNF, HaDeS will be upgraded with finer spatial granularity and a higher channel count to improve resolution. 
During low-intensity alignment scans, it can be inserted into the beamline to assess residual hadrons after the target, horn, and baffle components are installed. With this technology, beam-based alignments and failure diagnoses are possible with high precision.
The Muon Monitor System (MuMS), located downstream of the Hadron Absorber in three shielded alcoves, correlates proton beam properties with neutrino production. Each MuMS station consists of a $9 \times 9$ array of parallel-plate ionization chambers.
Each station provides 2D charge maps for each spill, revealing the transverse distribution of muons. These distributions encode valuable information about pion focusing, decay kinematics, and the effectiveness of the horn current settings. Because pions and muons share the same parentage with neutrinos, monitoring the muon beam allows inference of fluctuations in the resulting neutrino flux.
To study these relationships, we use the \texttt{g4lbnf} simulation package, which models the full LBNF beamline geometry. 
Simulations employ a Gaussian beam profile with $\sigma = 0.27$~cm and track pion decays and resulting muons through the beamline. In the nominal geometry, Alcove 1 receives muons from lower-energy pions emitted at wide angles, while Alcoves 2 and 3 capture muons more forward-going and higher-energy. 
This natural energy separation across the MuMS stations provides a handle for diagnosing beam focusing and energy spectra shifts. 
\begin{figure}[htbp]
\centering
\includegraphics[width=0.7\columnwidth]{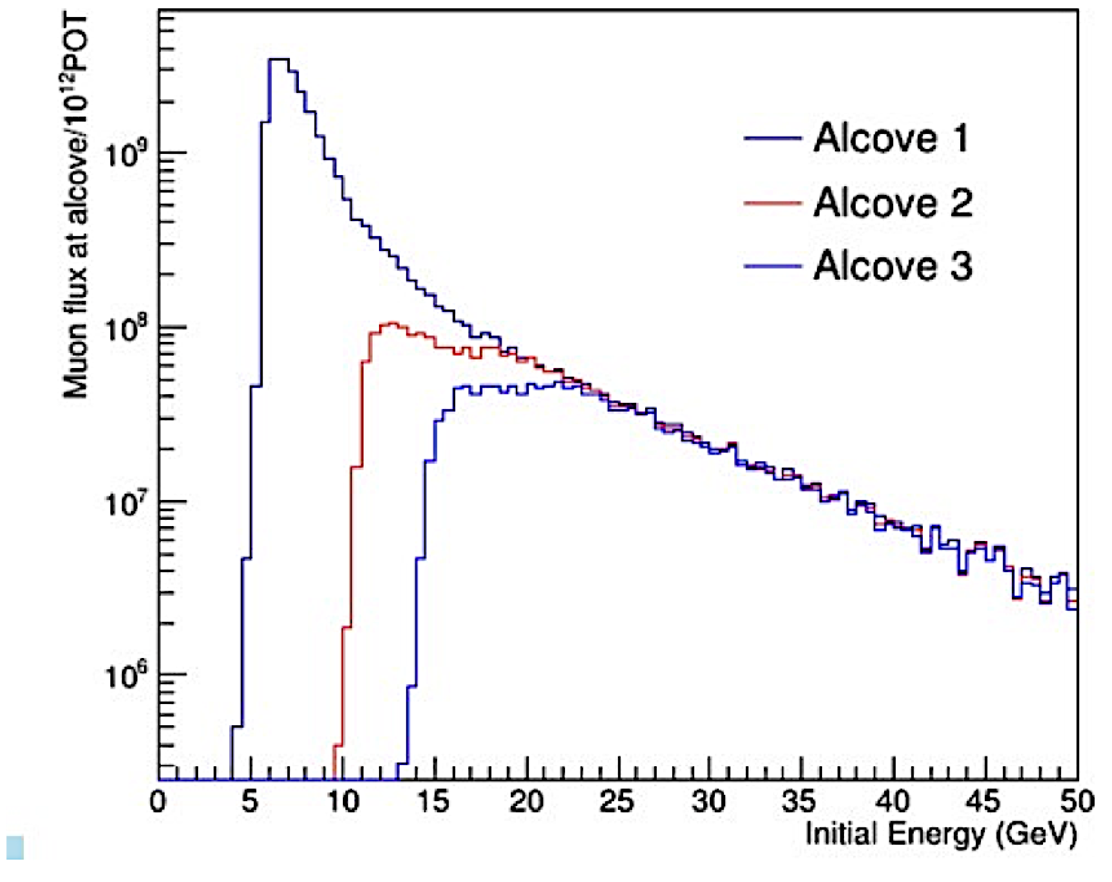} 
\caption{Simulated muon flux distributions at each MuMS alcove from \texttt{g4lbnf}. Alcove 1 captures the broadest angular range. Muon flux and centroid shifts provide indirect evidence of pion focusing and target degradation.}
\label{fig:muonflux}
\end{figure}
Figure~\ref{fig:muonflux} illustrates the simulated muon flux distributions at each MuMS alcove.
The design of the MuMS enables the study of both beam position sensitivity and beam spot size dependence. Tighter beam profiles result in more localized muon distributions, making centroid shifts more sensitive to upstream beam perturbations. Moreover, the energy-selective positioning of the alcoves provides an opportunity to decouple overlapping beam effects.

\begin{figure}[htbp]
\centering
\includegraphics[width=0.95\columnwidth]{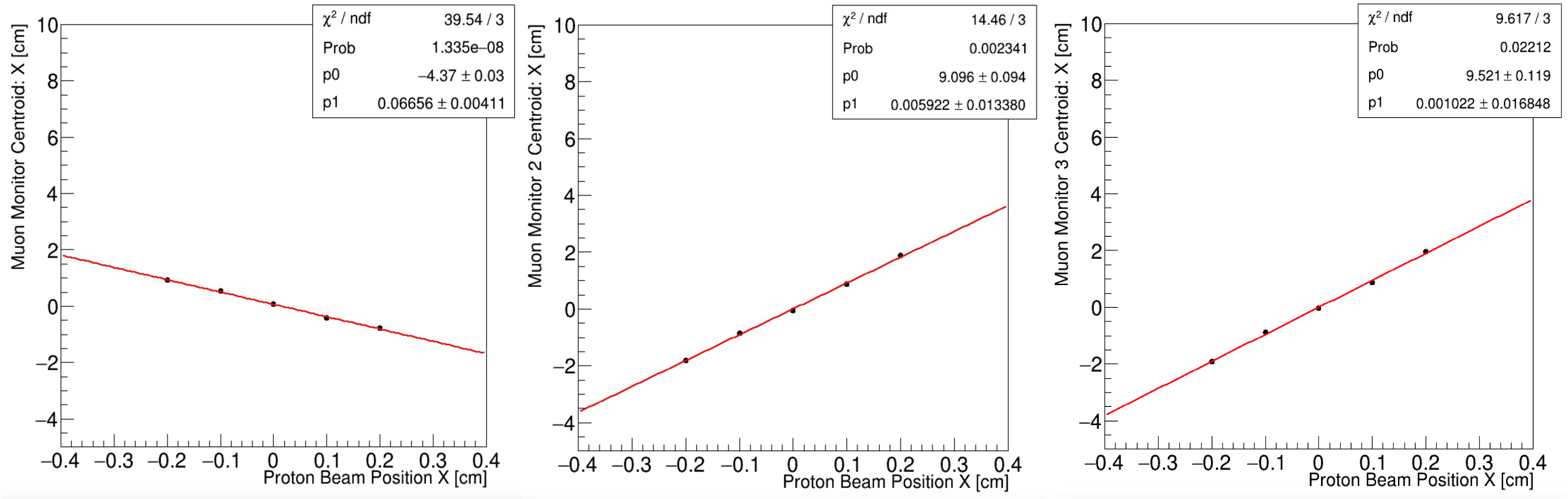} 
\caption{Linear fit of muon centroid position versus proton beam offset for MuMS1--3 (from left to right). Each slope reflects different energy sensitivity and focusing behavior.}
\label{fig:muoncentroids}
\end{figure}

Future multi-MW upgrades to LBNF will require muon monitors that are both radiation-hard and capable of resolving time-structured beam features. Candidate technologies include optical fiber-based Cherenkov detectors and fast ionization chambers with sub-nanosecond-scale timing. These could provide complementary time-resolved muon flux profiles, enabling more granular ML model inputs and real-time monitoring at unprecedented beam powers.
\begin{table*}[htbp]
    \centering
    \small 
    \caption{Summary of LBNF Beamline Instrumentation and Their ML Roles}
    \begin{tabular*}{\textwidth}{@{\extracolsep{\fill}} llll}
        \toprule
        \textbf{Instrument} & \textbf{Measurement} & \textbf{Location} & \textbf{ML Role} \\
        \midrule
        BPM / SEM & $(x, y)$ centroid, RMS width & Transport line to target & Provides beam position features for ML models \\
        Toroids / DCCT & Spill intensity (Protons On Target) & Upstream of target & Used to normalize model inputs and predictions \\
        BLMs & Localized losses & Transport / target chase & Flags unusual loss patterns for anomaly models \\
        Horn Current & Focusing strength, jitter & Horns 1--3 & Tunable input to ML-based beam control \\
        MuMS & 2D charge maps & Alcoves 1--3 & Input for inferring beam shape and alignment \\
        Hadron Monitor & Hadron energy deposit & Upstream of absorber & Helps train models to assess focusing accuracy \\
        SAND / ND-LAr & $\nu$ interactions & Near Detector & Ground truth, cross-check \\
        Env. Sensors & Temp, RF, humidity & Beamline Hall & Provides labeled data for model validation \\
        \bottomrule
    \end{tabular*}
    \label{tab:instrumentation}
\end{table*}
The response of each MuMS station varies with both proton beam alignment and horn settings as shown in Fig.~\ref{fig:muoncentroids}.
This figure shows that a linear correlation exists between the horizontal displacement of the proton beam on target and the centroid position of the muon beam in each alcove. The MuMS1 slope is negative, consistent with over-focusing of low-energy muons. In MuMS2 and MuMS3, the slopes are positive, suggesting that muons with low and medium energies are under-focused.

This redundancy in slope between MuMS2 and MuMS3 implies that both currently probe similar energy muons and may not independently improve diagnostics. Adjusting the energy thresholds or shielding of MuMS3 could restore sensitivity and make it complementary to MuMS2.
Overall, the MuMS and HaDeS systems form the backbone of real-time, spill-by-spill inference of beam parameters. As downstream detectors are limited to statistical flux measurements, these upstream systems enable rapid detection of anomalies and pave the way for feedback-based beam control in future ML frameworks.
Table~\ref{tab:instrumentation} summarizes the key LBNF beamline instruments and outlines their role in machine learning-based beam diagnostics. 
In addition to providing complementary measurements across the beamline, these devices also provide data for data-driven models. 
The integration makes it easier to detect anomalies, fine-tune beams, and infer pion phase spaces, which reduces systematic uncertainties in neutrino flux predictions.

\subsection{Spill-by-Spill Beam Inference and Integration with SAND}

While the MuMS system provides spill-resolved information on the beam's tertiary muon profile, it does not directly observe neutrino interactions. For this, the System for On-Axis Neutrino Detection (SAND)~\cite{battisti2019thesis} in the DUNE Near Detector complex plays a vital complementary role. SAND consists of a straw-tube tracker (STT), a liquid argon (GRAIN) module, and an electromagnetic calorimeter (ECAL) within a 0.6~T solenoid. It offers detailed event-by-event reconstruction of neutrino energy and flavor, allowing high-statistics flux characterization with $\sim$80 events per spill in neutrino mode~\cite{Sala2019DUNE}.

However, SAND is not capable of providing instantaneous beam feedback due to its need to accumulate sufficient statistics. In contrast, MuMS captures every spill and serves as a fast, indirect probe of beam alignment, target condition, and horn current drift. Machine learning models trained on simulated correlations between proton beam offsets, horn parameters, and MuMS charge maps can infer these upstream conditions in real time.

\begin{figure}[htbp]
    \centering
    \includegraphics[width=0.95\columnwidth]{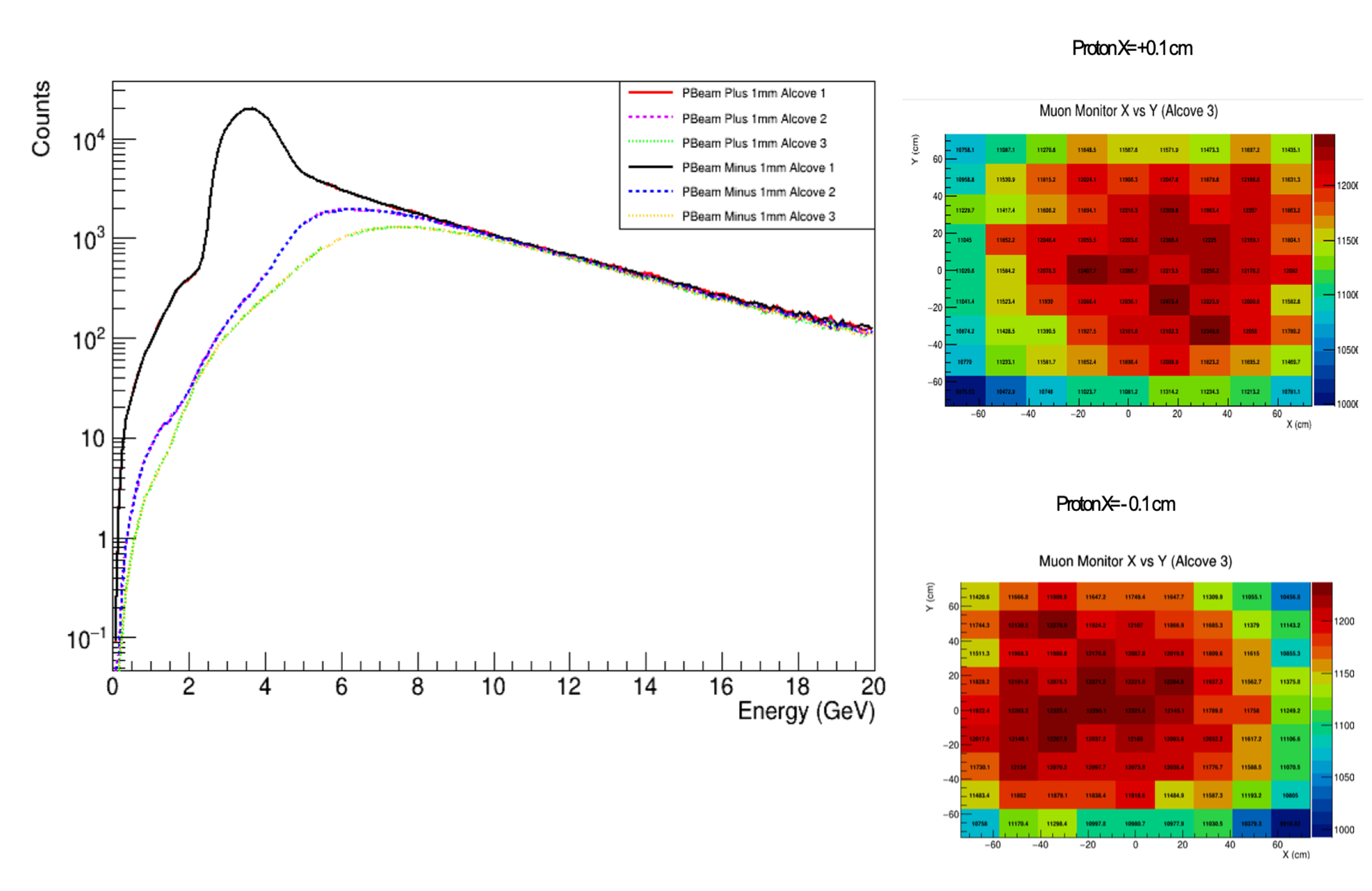} 
    \caption{Left: Neutrino energy spectra from G4LBNF simulation across MuMS alcoves. Right: Muon monitor 2D charge maps in Alcove 3 for $+1$~mm and $-1$~mm proton beam offsets. These spatial changes correlate with flux shifts observable in SAND.}
    \label{fig:sand_mums}
\end{figure}

Figure~\ref{fig:sand_mums} shows how spatial changes in muon distributions for positive and negative beam offsets manifest in energy spectra observable in the Near Detector. MuMS exhibits sensitivity to both high-energy ($\sim$4~GeV) and lower-energy ($\sim$2~GeV) neutrinos, though with reduced efficiency at lower energies, as well as the ability to detect changes in the proton beam position that may be invisible to the Near Detector.

\section{Machine Learning and Simulation Alignment}

Through the US--Japan collaboration, we have gained access to beamline data from the T2K experiment, which features a 7\(\times\)7 muon monitor array~\cite{Matsuoka_2010}. These detectors measure charge deposition from muons on a spill-by-spill basis, providing a spatially resolved signal proportional to the underlying proton beam properties. From these sensor arrays, four key beam parameters are extracted using Gaussian fits: beam center \((x_0, y_0)\), width \((\sigma_x, \sigma_y)\), and occasionally beam angle and horn current.  These are extracted with \(\sim3\) mm resolution and directional uncertainty of \(\sim0.2\) mrad.

T2K uses two types of sensor arrays: PIN photodiodes for fast, high-precision charge measurement and ionization chambers for radiation-tolerant operation. These are mounted on a movable stage, allowing alternating measurements for systematics studies. Sensor outputs are integrated over the beam spill and normalized to a precise current transformer, eliminating spill-to-spill variations in beam intensity and focusing model learning on beam shape and directionality.

We frame this as a supervised learning task: given the 49-pixel sensor map as input, can an ML model accurately reconstruct the corresponding beam parameters. To this end, we developed and trained both linear regression models and Fully Connected Neural Networks (FCNNs).

The linear regression model, implemented using TMVA, provides interpretability and fast training. Using raw MuMon signals as inputs, it predicts position, angle, width, and horn current. Weight heatmaps reveal sensor influence patterns, aligning with physical expectations (e.g., horn current weight concentrating at the edges due to high-energy muons). While effective across nominal ranges, this model degrades under horn current scans due to non-linear effects, showing residual bias near parameter extremes.

To overcome these limitations, we trained a more expressive FCNN model. The network architecture is 49\,$\rightarrow$\,7\,+\,1\,$\rightarrow$\,7\,+\,1\,$\rightarrow$\,7, also called the "777 network." This FCNN was trained on the same 50-shot dataset (\(\sim\)1k train, \(\sim\)1.9k test) using 150 epochs, a batch size of 32, a learning rate of 0.005, and a dropout of 0.1. Hyperparameter tuning was performed using Optuna. The model captures nonlinear correlations and generalizes well to beam conditions unseen during training, accurately predicting beam parameters during transitions and nonlinear scans.

\begin{figure}[htbp]
    \centering
    \includegraphics[width=0.45\textwidth]{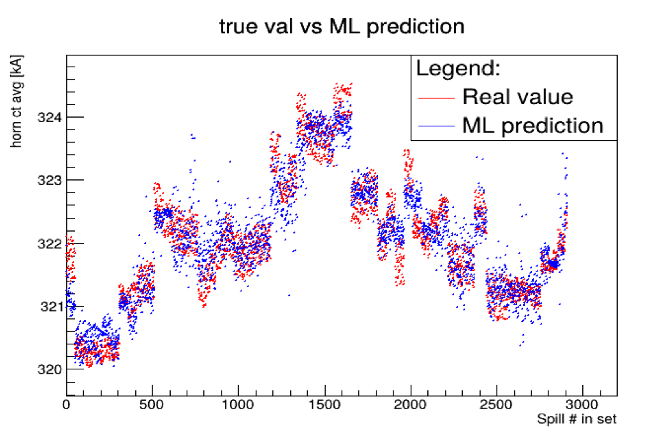}
    \hfill
    \includegraphics[width=0.45\textwidth]{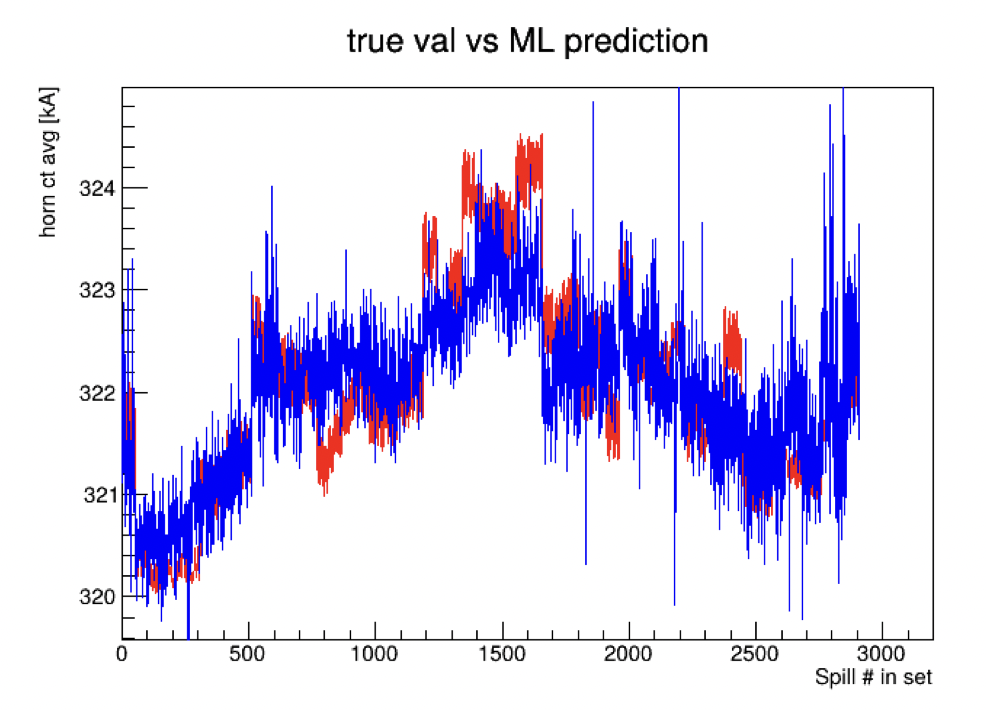}
    \caption{Time series prediction of horn current from T2K muon monitor data across 2900 spills. 
Top: FCNN model output (blue) tracks the true horn current (red) closely, including during sharp transitions. 
Bottom: Linear regression shows noticeable lag and deviation at extreme values, especially during non-linear horn current scans. 
Plots by Ian Heitkamp, adapted from~\cite{heitkamp_indico}.}

\label{fig:horn_prediction_comparison}
\end{figure}
\begin{figure}[htbp]
    \centering
     \includegraphics[width=0.75\columnwidth]{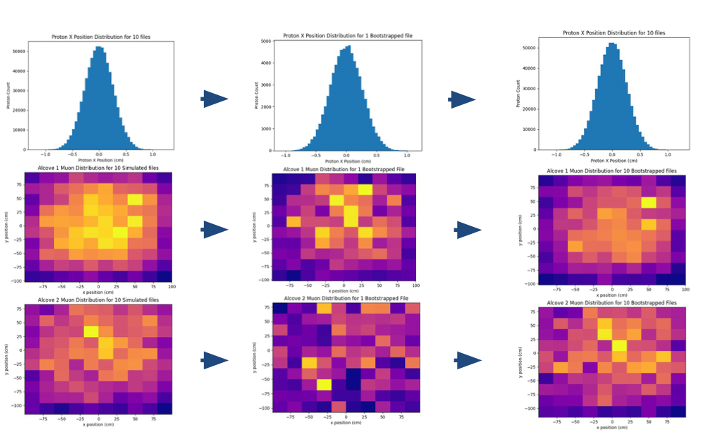} 
    \caption{Bootstrapped training strategy. Simulated MuMS data is perturbed within physical bounds to enhance model generalization without rerunning full simulations.}
    \label{fig:bootstrapping_flow}
\end{figure}
\begin{figure}[htbp]
    \centering
    \includegraphics[width=0.95\columnwidth]{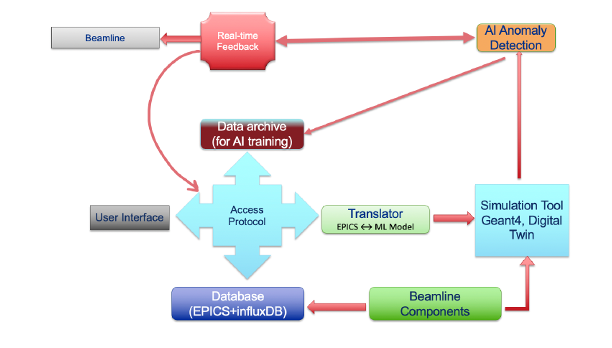} 
    \caption{Proposed ML control architecture for LBNF. Real-time MuMS data feeds anomaly detection, inference, and beam feedback via Digital Twin + RL agents.}
    \label{fig:ml_control_architecture}
\end{figure}

Building on these results, we propose an expanded ML-driven framework for LBNF, which integrates real-time beam inference, anomaly detection, and feedback control. At its core will be a physics-informed Digital Twin---a fast Geant4-based simulation of the beamline synchronized with live MuMS data. This will enable predictive modeling and ``what-if'' testing in under two spills. A Reinforcement Learning (RL) agent will explore horn currents and steering settings offline and use FCNN outputs as state vectors. It will initially be validated in shadow mode before the transition to active control.
A bootstrapping technique~\ref{fig:bootstrapping_flow} has been used to enhance statistical power when working with limited simulation data in order to improve model accuracy. Bayesian optimization is still in its exploratory stages, but it offers a promising method for selecting simulations for sampling. 
These innovations collectively aim to transition from passive monitoring to intelligent, feedback-driven operations.
The proposed architecture is shown in Fig.~\ref{fig:ml_control_architecture}.
The complete pipeline structure is summarized in Table~\ref{tab:ml_lbnf_plan} maps ML components to data sources and deployed models. We use FCNNs to infer parameters, auto-encoders to detect anomalies, and RL agents to control closed loops.
The whole system communicates with the Experimental Physics and Industrial Control System (EPICS) through a translator.

\begin{table*}[htbp]
    \centering
    \caption{End-to-end ML Control Strategy for LBNF Beamline}
    \begin{tabular*}{\textwidth}{@{\extracolsep{\fill}} p{3.5cm}p{5.5cm}p{3.5cm}p{3.5cm}}
        \toprule
        \textbf{Component} & \textbf{Summary} & \textbf{Data Used} & \textbf{Model Used} \\
        \midrule
        Why move from monitoring to control? & Already watch beam with ML; next step is automated feedback & Real MuMS data & FCNN/Autoencoder for real-time monitoring \\
        Reinforcement Learning (RL) & Simulated beamline lets RL test horn/steering settings with reward feedback & Simulated data + real data for shadow validation & RL Agent + FCNN outputs as state input \\
        Physics-informed Digital Twin & Mini-Geant4 synced to live MuMS data enables fast ``what-if'' testing & Simulated + real MuMS data & Shared by FCNN and RL as physical model \\
        Whole data loop & Beam \(\rightarrow\) MuMS \(\rightarrow\) model \(\rightarrow\) control suggestion \(\rightarrow\) beam & MuMS real data; RL trained on sim, fine-tuned on real & FCNN extracts (x0, \(\sigma_x\), etc.), AE flags anomalies, RL drives Twin \\
        Operators & Interface for control and what-if forecasting; transition from monitoring to feedback & Simulated + real data & RL and AE suggest real-time decisions \\
        \bottomrule
    \end{tabular*}
    \label{tab:ml_lbnf_plan}
\end{table*}

This combined FCNN+AE+RL pipeline enables spill-level inference of beam parameters, robust anomaly detection, and automated beam tuning---transforming monitoring systems into intelligent beam control agents for LBNF and future high-power neutrino programs.

\section{Advanced Instrumentation Development}

The transition to megawatt-class proton beams imposes new demands on downstream diagnostics. Radiation tolerance, high occupancy, and nonlinear beam dynamics challenge the performance of existing instrumentation. The current Muon Monitor System (MuMS), composed of ionization chamber arrays in Alcoves 1--3, has been robust and radiation-hard, providing reliable 2D charge maps per spill. However, it offers only integrated charge over each spill and limited temporal information, which restricts its utility in resolving fast, complex beamline behavior.

New radiation-hard technologies allow us to preserve functionality while also expanding diagnostics. Specifically, determining the time-of-flight (TOF) of muons between decay volume and MuMS presents a previously unexplored, but increasingly valuable, technique. 
There are two reasons:

\begin{enumerate}
    \item With upgraded detector technologies already under consideration for radiation hardness and fine segmentation, adding timing capability incurs minimal additional cost but yields significant gains in information.
    \item Timing enables event-by-event correlation between muon arrival time and momentum, allowing us to directly validate Geant4-based simulation predictions and monitor beam optics performance in new ways.
\end{enumerate}

\begin{figure}[ht]
  \centering
  \includegraphics[width=0.75\columnwidth]{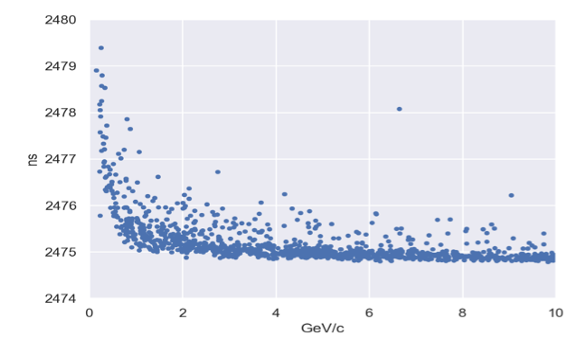}
  \caption{Simulated time-of-flight versus muon momentum at Muon Monitor 1. The timing difference between 6.7 and 7~GeV/$c$ muons is approximately 26~ps at Alcove~1.}
  \label{fig:tof_sim}
\end{figure}

Figure~\ref{fig:tof_sim} shows a simulation of TOF as a function of muon momentum at Muon Monitor~1. Even for a narrow momentum range (6.7--7~GeV/$c$), the expected arrival time difference is on the order of 26~ps—well within reach of modern picosecond-scale detectors. Such measurements can reveal chromatic effects from magnetic horn focusing, changes in target geometry, or off-nominal optics configurations. Moreover, the observed time structure varies across pixels, so we get an extra layer of resolution.

To implement this capability, several detector technologies are being evaluated:
\begin{itemize}
  \item \textbf{Low-Gain Avalanche Diodes (LGADs)} — thin silicon sensors with internal gain and timing resolution of 30--50~ps, already deployed in collider environments;
  \item \textbf{Cherenkov-based detectors} — compact radiators with picosecond-scale response, coupled to Silicon Photomultipliers (SiPMs) or Microchannel Plate PMTs (MCP-PMTs);
  \item \textbf{MCP-PMTs} — vacuum photodetectors with demonstrated <50~ps resolution and high radiation tolerance, used in test beams and timing layers.
\end{itemize}

Incorporating timing into MuMS would enhance the spatial-temporal resolution of beam monitoring and enable new machine learning (ML) workflows. Time-resolved charge maps allow models to disentangle momentum-dependent effects, detect subtle anomalies, and track variations over time. 
Also, TOF measurements can be used to compare real and synthetic data in the development of digital twins, allowing simulation validation.

\section{Toward Intelligent Beam Control and Systematics Reduction}
The new instrumentation enables higher-resolution, potentially time-tagged muon data, so we can do precision inference and control. 
We can now use muon monitors to actively characterize and steer the beam instead of treating them as passive diagnostics. 

\begin{itemize}
    \item MuMS observes spill-by-spill muon distributions.
    \item ML models infer beam center, width, angle, and horn settings.
    \item Predictions are attached as metadata to each spill.
    \item Later, SAND measurements can be reweighted using this metadata to reduce systematic uncertainties.
\end{itemize}

In practice, this strategy may reduce flux systematics from 2-3\% to under 1\%, accelerating sensitivity to CP violation by several years. 

\section{Results and Outlook}
For multi-MW LBNF operations, enhanced radiation-hard muon monitors with nanosecond timing will further improve these capabilities. Fast Cherenkov detectors or ionization chambers with picosecond-scale electronics could allow event-by-event matching between muons and neutrinos, unlocking next-generation feedback and control for neutrino experiments.

\section{CONCLUSION}
We have presented the concept of a machine-learning-based framework that transforms conventional muon monitoring into a high-resolution, inference-capable diagnostic system for future high-power neutrino beams. 
By leveraging spill-by-spill charge maps from the T2K Muon Monitor System and training both linear and fully connected neural networks, we demonstrated precise reconstruction of beam parameters including center, width, angle, and horn current—even under nonlinear conditions such as horn scans.
These techniques were further embedded into a physics-informed control architecture, featuring a real-time digital twin and reinforcement learning agents for beam tuning. 
This closed-loop design enables rapid detection of drifts, adaptive control, and substantial reduction of beam-related flux systematics—from the current 2–3\% level to below 1\%. 
These reductions result in faster convergence toward CP violation sensitivity and improved operational resilience when applied to LBNF and DUNE.
As new timing technologies emerge, muon monitors could be enhanced with time-of-flight information to reconstruct phase space event by event.
This combination of fast sensors, intelligent models, and hybrid simulation lays the groundwork for the next generation of accelerator-based neutrino experiments—where real-time intelligence is embedded in the beamline itself.

\section{ACKNOWLEDGEMENTS}
This work was produced by Fermi Forward Discovery Group, LLC under Contract No. 89243024CSC000002 with the U.S. Department of Energy, Office of Science, Office of High Energy Physics. Publisher acknowledges the U.S. Government license to provide public access under the DOE Public Access Plan\url{https://www.energy.gov/doe-public-access-plan}.

%
%
\ifboolexpr{bool{jacowbiblatex}}%
	{\printbibliography}%

\begin{thebibliography}{9} 
	\bibitem{papadimitriou:ipac16-tupmr025}
V. Papadimitriou \emph{et al.}, “Design of the LBNF Beamline,” in \emph{Proc. IPAC'16}, Busan, Korea, Jun. 2016, pp. 1291–1293, doi:10.18429/JACoW-IPAC2016-TUPMR025. Available: \url{https://jacow.org/ipac2016/papers/TUPMR025.pdf}
        \bibitem{papadimitriou:ipac18-tupaf075}
V. Papadimitriou \emph{et al.}, “Design Status of the LBNF/DUNE Beamline,” in \emph{Proc. IPAC'18}, Vancouver, Canada, Apr.–May 2018, pp. 902–904, doi:10.18429/JACoW-IPAC2018-TUPAF075. Available: \url{http://accelconf.web.cern.ch/ipac2018/papers/TUPAF075.pdf}
        \bibitem{q6l6-wywy}
K.~Yonehara, S.~Ganguly, D.~A.~Wickremasinghe, P.~Snopok, and Y.~Yu,
``Precision beam diagnostics at the NuMI facility using muon monitor observations,''
\textit{Phys. Rev. Accel. Beams}, American Physical Society, Jul. 2025.
\href{https://link.aps.org/doi/10.1103/q6l6-wywy}{doi:10.1103/q6l6-wywy}.
        \bibitem{Zwaska_2006}
R.~Zwaska, M.~Bishai, S.~Childress, G.~Drake, C.~Escobar, P.~Gouffon, D.~A.~Harris, J.~Hylen, D.~Indurthy, G.~Koizumi, S.~Kopp, P.~Lucas, A.~Marchionni, A.~Para, Ž.~Pavlović, W.~Smart, R.~Talaga, and B.~Viren, 
``Beam-based alignment of the NuMI target station components at FNAL,'' 
\emph{Nucl. Instrum. Meth. A}, vol.~568, no.~2, pp.~548--560, Dec. 2006, 
doi:~\href{https://doi.org/10.1016/j.nima.2006.08.031}{10.1016/j.nima.2006.08.031}.
        \bibitem{battisti2019thesis} F. Battisti, Monitoring of the DUNE Long Baseline Neutrino Beam with the SAND Detector, Bachelor's thesis, University of Bologna, 2019.

        \bibitem{Sala2019DUNE} 
P. A. Sala, ``A Proposal to Enhance the DUNE Near-Detector Complex,'' DUNE-doc-13262-v7, March 2019. Available: \url{https://docs.dunescience.org/cgi-bin/sso/RetrieveFile?docid=13262&filename=A_Near_Detector_for_DUNE.pdf&version=7}.
        \bibitem{Matsuoka_2010}
K.~Matsuoka, A.~K.~Ichikawa, H.~Kubo, K.~Maeda, T.~Maruyama, C.~Matsumura, A.~Murakami, T.~Nakaya, K.~Nishikawa, T.~Ozaki, K.~Sakashita, K.~Suzuki, S.~Y.~Suzuki, K.~Tashiro, K.~Yamamoto, and M.~Yokoyama,
\newblock ``Design and performance of the muon monitor for the T2K neutrino oscillation experiment,''\\
\newblock {\em Nucl. Instrum. Methods Phys. Res. A}, vol.~624, no.~3, pp.~591--600, Dec. 2010,\\
\newblock \href{http://dx.doi.org/10.1016/j.nima.2010.09.074}{DOI: 10.1016/j.nima.2010.09.074}.
\bibitem{heitkamp_indico}
Ian Heitkamp and Esra Barlas, \textit{T2K MuMon ML Studies},\\
\url{https://indico.fnal.gov/event/67937/contributions/308681/attachments/185041/254628/T2K\%20Mumon\%20ML\%20Studies.pdf},\\
Accessed: July 2025.
        
\end{thebibliography}
	{%
	

}
\end{document}